\documentclass{emulateapj}
\usepackage{natbib}
\usepackage{amsmath}
\usepackage{graphicx}
\usepackage{color}
\usepackage{epsfig}
\usepackage{natbib}
\newcommand{\K}{{\rm K}}

\newcommand{\msun}{{\rm M}_\odot}

\newcommand{\mdot}{\dot{M}_{\rm BH}/\dot{M}_{\rm Edd}}
\newcommand{\msunyr}{M_\odot~{\rm yr}^{-1}}

\newcommand\lsim{\mathrel{\rlap{\lower4pt\hbox{\hskip1pt$\sim$}}
        \raise1pt\hbox{$<$}}}
\newcommand\gsim{\mathrel{\rlap{\lower4pt\hbox{\hskip1pt$\sim$}}
        \raise1pt\hbox{$>$}}}

\slugcomment{}

\shorttitle{A maximum mass of SMBHs}
\shortauthors{Kohei Inayoshi \& Zolt\'an Haiman}

\begin{document}

\title{Is there a maximum mass for black holes in galactic nuclei?}

\author{Kohei Inayoshi}
\author{Zolt\'an Haiman}
\affil{Department of Astronomy, Columbia University, 550 West 120th Street, New York, NY 10027, USA}
\email{KI: inayoshi@astro.columbia.edu -- Simons Society of Fellows}

\begin{abstract}
The largest observed supermassive black holes (SMBHs) have a mass of
$M_{\rm BH}\simeq 10^{10}~\msun$, nearly independent of redshift,
from the local ($z\simeq 0$) to the early ($z>6$) Universe.  We
suggest that the growth of SMBHs above a few $\times 10^{10}~\msun$ is
prevented by small-scale accretion physics, independent of the
properties of their host galaxies or of cosmology.  Growing more
massive BHs requires a gas supply rate from galactic scales onto a
nuclear region as high as $\gsim 10^3~\msunyr$.  At such a high
accretion rate, most of the gas converts to stars at large radii
($\sim 10-100$ pc), well before reaching the BH.  We adopt a simple
model \citep{TQM05} for a star-forming accretion disk, and find that
the accretion rate in the sub-pc nuclear region is reduced to the
smaller value of at most a few $\times~\msunyr$.  This prevents SMBHs from
growing above $\simeq 10^{11}~\msun$ in the age of the Universe.
Furthermore, once a SMBH reaches a sufficiently high mass, this rate
falls below the critical value at which the accretion flow becomes
advection dominated.  Once this transition occurs, BH feeding can be
suppressed by strong outflows and jets from hot gas near the BH.  We
find that the maximum SMBH mass, given by this transition, is
between $M_{\rm BH,max}\simeq (1-6) \times 10^{10}~\msun$,
depending primarily on the efficiency of angular momentum transfer
inside the galactic disk, and not on other properties of the host
galaxy.
\end{abstract}


\section{Introduction}

Most massive galaxies in the local Universe are inferred to host
supermassive black holes (SMBHs) with masses of $10^5-10^{10}~\msun$
at their centers.  The correlations observed between the masses
($M_{\rm BH}$) of the SMBHs and the velocity dispersion ($\sigma$) and
other bulk properties of their host galaxies suggest that they
co-evolved during their cosmic history \citep[e.g.][and references
  therein]{Kormendy_Ho_2013}.  The correlations could be caused by BH
feedback, which can suppress star formation and gas supply on galactic
scales
\citep[e.g.][]{Silk_Rees_1998,1999MNRAS.308L..39F,2003ApJ...596L..27K,2005ApJ...618..569M}.
These observations have also revealed a maximum SMBH mass of $\sim
10^{10}~\msun$, in the largest elliptical galaxies
\citep[e.g.][]{McConnell_2011}.

Observations of distant quasars, with redshift as high as $z\sim 7$,
have found that the SMBH masses fueling the brightest quasars are
similarly $\sim 10^{10}~\msun$
\citep[e.g.][]{Fan2001,2010AJ....139..906W,Mortlock_2011,Wu_2015}.
Intriguingly, this apparent maximum mass is nearly independent of
redshift \citep[e.g.][]{Netzer_2003,Vestergaard_2004,Marconi_2004,Trakhtenbrot_2012}.
Since the e-folding time for BH mass growth (at the fiducial
Eddington-limited accretion rate, with a 10\% radiative efficiency) is
$\sim 40$ Myr, much shorter than the cosmic age. Given sufficient
fuel, SMBHs could thus continue to grow, and reach masses well above
$\sim 10^{10}~\msun$ by $z\simeq 0$.  However, we do not see SMBHs
significantly above $\sim 10^{10}~\msun$ in the local Universe (or
indeed at intermediate redshift).

Naively, the near-constant value of the maximum SMBH mass with
redshift is therefore surprising.  It is tempting to attribute this
observation to the same galactic-scale feedback that ties SMBH masses
to their host galaxies.  The maximum masses of galaxies in a fixed
comoving volume are determined by the physics of cooling and galactic
feedback processes, but in general, they should increase as galaxies
are assembled over time.  However, local surveys probe smaller
comoving volumes than high-$z$ surveys, and can miss the rarest, most
massive galaxies. In principle, this could coincidentally lead to a
maximum galaxy mass that stays roughly constant with redshift.  In
practice, this explanation requires the $M_{\rm BH}-\sigma$
correlation to evolve \citep{Netzer_2003,NatarajanTreister2009}, and
also the quasar luminosity function to steepen at the bright end
\citep{NatarajanTreister2009}.

Here we pursue a possible alternative interpretation. Namely, the
observations suggest that SMBHs stopped growing at near-Eddington
short after $z\simeq 5$, once they reached $\sim 10^{10}~\msun$
\citep{Trakhtenbrot+2011}.
On the other hand, galaxies do not likewise stop their growth at this
early epoch: the most massive ellipticals are believed to have
assembled at $z\simeq 1-2$
\citep[e.g.][]{2003AJ....125.1882B,2005ApJ...621..673T}.  This
motivates us to hypothesize that there is a limiting mass, determined
by small-scale physical processes, independent of galaxy evolution,
star formation history, or background cosmology.  In this paper, we
discuss such a ``microphysical'' scenario, limiting the growth of
SMBHs to a few$\times 10^{10}~\msun$: disks with the high accretion
rates required to produce more massive SMBHs fragment into stars. The
small residual fraction of gas that trickles to the inner region is
unable to form a standard geometrically thin accretion disk and to
accrete onto the BH, and is instead expelled in winds or
jets.\footnote{As this paper was being completed, we became aware of a
  recent preprint proposing a similar idea \citep{King_2016}. We
  discuss the similarities and differences between the two works in
  \S\ref{sec:discussion} below.}

The rest of this paper is organized as follows.
In~\S\ref{sec:mass_limit}, we discuss the model for star-forming
accretion disks, and the implied maximum SMBH mass.
In~\S\ref{sec:BH_lum}, we show that our results can explain the
observed $M_{\rm BH}-L_{\rm bol}$ relation for most AGN/QSOs, as well
as the maximum SMBH mass.  In~\S\ref{sec:discussion}, we discuss
possible caveats, and in \S\ref{sec:conclusions} we summarize our
conclusions.  Throughout this paper, we define the Eddington accretion
rate as $\dot{M}_{\rm Edd}\equiv 10~L_{\rm Edd}/c^2=230~\msunyr
(M_{\rm BH}/10^{10}~\msun)$.

\section{Limit on SMBH growth via an accretion disk}
\label{sec:mass_limit}

\subsection{Star-forming accretion disks}\label{sec:TQM}

We here consider a model for a star-forming accretion disk around a
SMBH with $M_{\rm BH}\sim 10^{8-11}~\msun$ based on \citet[][hereafter
  TQM05]{TQM05}.  In this model, the gas fueling rate from galactic
scales ($\ga 100$ pc) to the nuclear region ($\la 1$ pc) is estimated
self-consistently, including gas depletion due to star-formation.
Because of star-formation, the central BH is fed at a rate of 
$<\dot{M}_{\rm Edd}$ and thus the BH growth is limited.
This is consistent with most observed AGNs/QSOs,
whose Eddington ratios are inferred to be modest (e.g. $L/L_{\rm Edd}
\sim 0.16$ for $0.35<z<2.25$ and $L/L_{\rm Edd} \sim 0.43$ for $z >
4$; \citealt{Shen_2011,DeRosa_2011}).  A few exceptionally bright QSOs
at higher redshift are believed to accrete more rapidly, at or even
somewhat above $\sim \dot{M}_{\rm Edd}$.
In the high-rate case, fragmentation of a nuclear disk suppresses the BH feeding
(see discussion in \S\ref{sec:discussion} and \citealt{King_2016}).

The TQM05 model assumes that radiation pressure from stars forming in
the disk supports the gas against gravity in the vertical direction,
and keeps the disk marginally stable; the Toomre parameter is then
\begin{equation}
Q\simeq \frac{c_{\rm s}\Omega}{\pi G \Sigma_{\rm g}} \simeq 1,
\label{eq:Q}
\end{equation}
where $c_s$ is the sound speed, $\Sigma_{\rm g}$ is the gas surface
density, and $\Omega$ is the orbital frequency, given by
\begin{equation}
\Omega=\left(\frac{GM_{\rm BH}}{r^3}+\frac{2\sigma ^2}{r^2} \right)^{1/2}.
\label{eq:omega}
\end{equation}
Here $\sigma$ is the velocity dispersion characterizing the
gravitational potential on galactic scales.  From the continuity
equation, the surface density is given by
\begin{align}
\Sigma_{\rm g}=\frac{\dot{M}}{2\pi r v_r}=\frac{\dot{M}}{2\pi r m c_{\rm s}}
\label{eq:Sigma}
\end{align}
where $\dot{M}$ is the gas accretion rate through a radius of $r$,
$v_r$ is the radial velocity and $m~(=v_r/c_{\rm s})$ is the radial
Mach number.  Note that the viscosity in this model is specified by
assuming a constant value of $m$ (see below), instead of the
$\alpha$-prescription \citep{Shakura_Sunyaev_1973}.

The disk is supported vertically by both thermal gas pressure ($p_{\rm
  gas}=\rho k_{\rm B}T/m_{\rm p}$) and radiation pressure due to stars
in the disk, where $\rho=\Sigma_{\rm g}/(2h)$ is the gas density,
$h=c_{\rm s}/\Omega$ is the pressure scale height and $T$ is the gas
temperature.  The radiation pressure is given by
\begin{equation}
p_{\rm rad}= \epsilon \dot{\Sigma}_\ast c \left(\frac{\tau}{2}+\xi \right),
\label{eq:prad}
\end{equation}
where $\tau=\kappa \Sigma_{\rm g}/2$ is the optical depth, $\kappa$ is
the dust opacity \citep{Semenov_2003}, $\dot{\Sigma}_\ast$ is the
star-formation rate per unit disk surface area 
and $\epsilon$ is the matter-radiation conversion efficiency, which depends on the mass function of stars.
The first term on the right-hand side of Eq. (\ref{eq:prad}) is the radiation pressure on
dust grains in the optically thick limit ($\tau\gg1$), and the second
term represents stellar UV radiation pressure and turbulent support by
supernovae in optically thin limit ($\tau \ll 1$), which is characterized by the non-dimensional value of $\xi$.
Energy balance between cooling and heating is given by
\begin{equation}
\sigma_{\rm SB}T_{\rm eff}^4 = \frac{1}{2}\epsilon \dot{\Sigma}_\ast c^2 + \frac{3}{8\pi }\dot{M}\Omega^2,
\label{eq:energy}
\end{equation}
where the effective temperature $T_{\rm eff}$ is given by
\begin{equation}
T^4=\frac{3}{4}T_{\rm eff}^4 \left(\tau + \frac{2}{3\tau} +\frac{4}{3}\right).
\label{eq:Teff}
\end{equation}
In this disk, a fraction of gas forms stars at a rate of
$\dot{\Sigma}_\ast(r)$ and the gas accretion rate decreases inward,
given by
\begin{equation}
\dot{M}(r)=\dot{M}_{\rm out} -\int^r_{R_{\rm out}}2\pi r \dot{\Sigma}_\ast dr.
\label{eq:Mdot}
\end{equation}

Eqs. (\ref{eq:Q})--(\ref{eq:Mdot}) determine the radial profiles of
all physical quantities, once the five parameters $M_{\rm BH}$,
$\sigma$, $m$, $\epsilon$, $\xi$ and the outer boundary conditions of
the accretion rate $\dot{M}_{\rm out}$ at the radius $R_{\rm out}$ are
chosen.  As our fiducial model, we set $\epsilon=10^{-3}$, appropriate
for a Salpeter initial mass function (IMF) with $1-100~\msun$, and $\xi=1$,
appropriate when turbulent support by supernovae is negligible, as in
a high-$\dot{M}$ (or $\rho$) disk.  The velocity dispersion is set to
$\sigma=400~{\rm km~s}^{-1}$, motivated by the empirical correlation
between BH mass and $\sigma$ of its host galaxy for $M_{\rm BH}\sim
10^{10}~\msun$
\citep[e.g.][]{2002ApJ...574..740T,2009ApJ...698..198G}.
Note that the dependence of our results on the choice of $\sigma$ is very weak
because the stellar gravitational potential is subdominant at
$r\la GM_{\rm BH}/(2\sigma^2)\simeq 140~{\rm pc}
~(M_{\rm BH}/10^{10}~\msun)(\sigma/400~{\rm km~s}^{-1})^{-2}$,
where the BH feeding rate is determined.

We consider a very high accretion rate of $\dot{M}_{\rm out}=10^3~\msunyr$ at the boundary, 
in order to give a conservative estimate on the maximum BH mass.  
From cosmological simulations \citep[e.g.][]{Genel_2009,Fakhouri_2010}, the maximum gas accretion
rate onto a dark matter halo is estimated as $\la 10^3~\msunyr (M_{\rm halo}/10^{12}~\msun)^{1.1}[(1+z)/7]^{5/2}$, 
where $M_{\rm h}$ is the halo mass.  This could be exceeded only for brief periods during major
merger events in the early Universe \citep{Mayer+2015}, and for
$M_{\rm h}\ga 10^{12}~\msun$, gas heating by a virial shock prevents
cold gas supply because of inefficient radiative cooling
\citep{Birnboim_2003,Dekel_2006}.

The nature of the angular momentum transfer, allowing gas to flow from
large scales to the inner sub-pc regions remains uncertain.  Following
TQM05, we assume the transfer is provided by global density waves, and
also that the radial Mach number $m$ is constant (independent of radius).  
Since our results depend on the choice of $m$, we here
consider three cases of $m=0.05$, $0.1$ and $0.2$.  These values are
motivated by analytical arguments yielding the limit $m \la 0.2$
\citep{2003MNRAS.339..937G}.
Note that in terms of the standard $\alpha$-prescription 
as a model for the disk viscosity \citep{Shakura_Sunyaev_1973},
the viscous parameter is related to the Mach number as $\alpha = m(2r/3h)$.

\begin{figure}
\centering
\includegraphics[width=80mm]{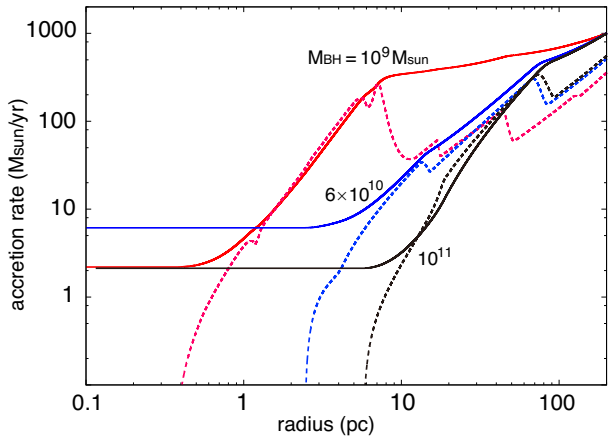}
\caption{Gas accretion rate (solid) and star formation rate (dashed)
  in a star-forming accretion disk.  The curves correspond to SMBH
  masses of $M_{\rm BH}=10^9$ (red), $6\times 10^{10}$ (blue), and
  $10^{11}~\msun$ (black).  The accretion rate at the outer boundary
  ($R_{\rm out}=200$ pc) is set to $\dot{M}_{\rm out}=10^3~\msunyr$.
  In each case, the accretion rate in the inner region ($\la 1$ pc)
  approaches a constant value, which is much smaller than
  $\dot{M}_{\rm out}$ because of star-formation at larger radii.}
\label{fig:TQM}
\vspace{2\baselineskip}
\end{figure}

Fig. \ref{fig:TQM} shows radial profiles of the gas accretion rate
(solid) and star formation rate (dashed) for three different BH masses
and for $m=0.1$.  For the lowest BH mass ($M_{\rm BH}=10^9~\msun$; in
red), star formation is inefficient at $r\ga 10$ pc, and the accretion
rate remains close to its value at the outer boundary.  At $r\lsim10$
pc, vigorous star-formation depletes most of the gas, and the
accretion rate rapidly decreases inward.  In this domain, the disk
temperature reaches the dust sublimation temperature ($T_{\rm
  dust,sub}\simeq 10^3$ K), above which the dust opacity drops
rapidly.  Since $\dot{\Sigma}_\ast \propto \Sigma_{\rm g}/\kappa$ in
the optically thick limit 
(see Eq. \ref{eq:3} in the Appendix), 
a higher star formation rate is required to
maintain the marginally-stable disk structure with $Q\simeq 1$
when the opacity decreases.
Note that since dust is composed of multiple species and each has a 
different sublimation temperature \citep{Semenov_2003},
the jagged radial profile for the star formation rate at $r\sim 10-100$ pc is caused by 
small drops of dust opacity at the corresponding sublimation temperature.
Within $r\lsim0.5$ pc, the disk becomes stable, star formation ceases,
and the gas accretion rate approaches a constant value.  
This accretion rate in the nuclear region does not depend on the value of
$\dot{M}_{\rm out}$, as long as $\dot{M}_{\rm out}> \dot{M}_{\rm crit} \simeq 280~\msunyr$ 
(see Eq. \ref{eq:12} in the Appendix),
and thus hardly on the model parameters of the star-forming disk except the radial Mach number 
$m$ (see \S\ref{sec:subpc}).
For higher SMBH masses, the accretion rate in the nuclear region
gradually increases.  However, for $M_{\rm BH}\ga 6\times
10^{10}~\msun$, the accretion rate at $<1$ pc decreases again, because
gas is depleted more efficiently due to star formation at larger radii
($r\sim 100$ pc), where evaporation of volatile organics decreases the
dust opacity moderately ($T\ga 400$ K) and 
the star formation rate increases in the optically thick star-burst disk
($\dot{\Sigma}_\ast \propto \Sigma_{\rm g}/\kappa$).

\subsection{Accretion in the sub-pc nuclear region}\label{sec:subpc}

\begin{figure}
\centering
\includegraphics[width=80mm]{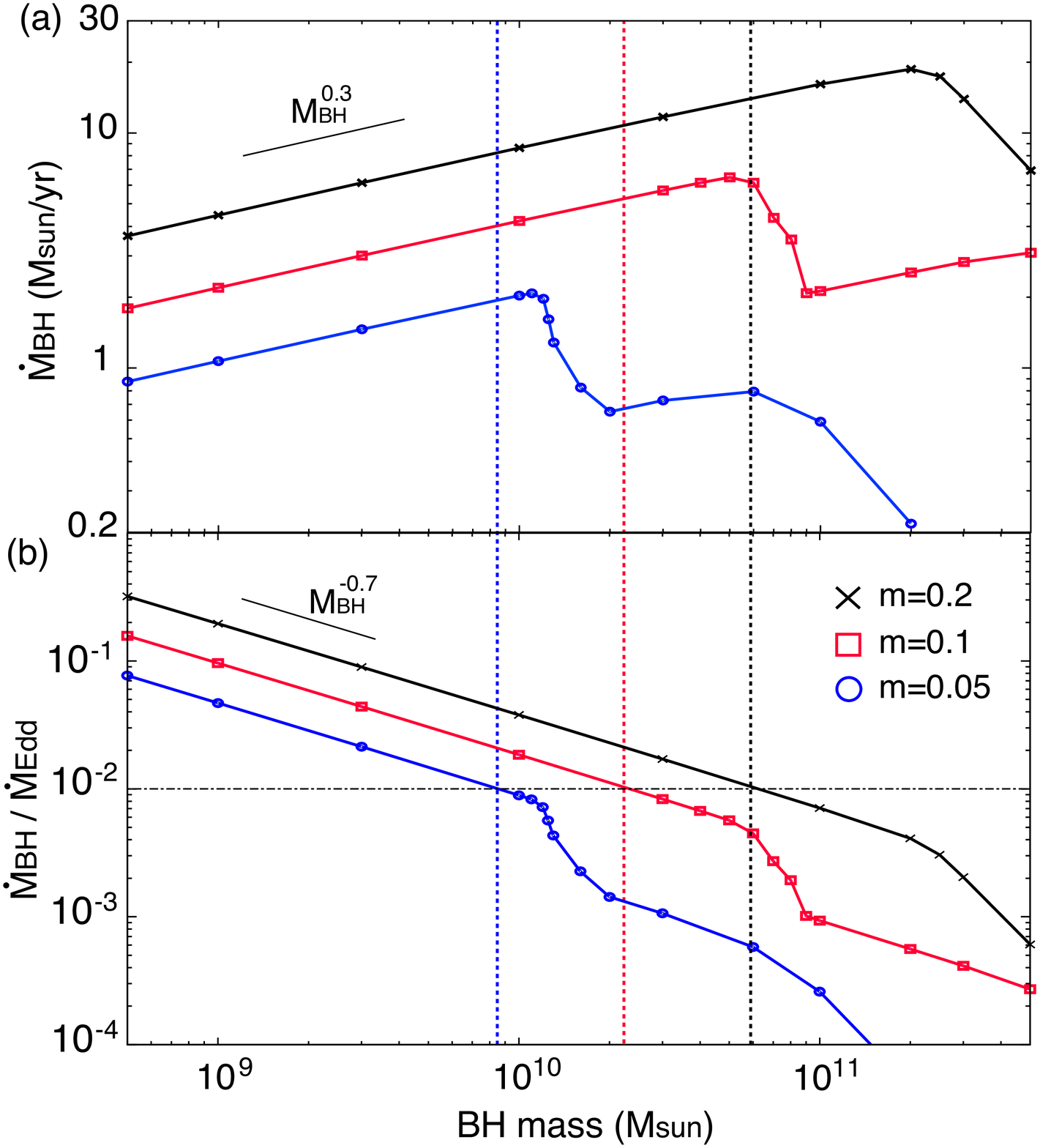}
\caption{Gas accretion rate into the nuclear region ($<1$ pc) as a
  function of SMBH mass, for three different angular momentum transfer
  efficiencies; $m=0.05$ (blue), $0.1$ (red) and $0.2$ (black).  The
  other parameters are the same as in Fig. \ref{fig:TQM}.  The
  horizontal line in the bottom panel marks $\mdot=10^{-2}$, below
  which a thin disk changes to an ADAF
  \citep{Narayan_McClintock_2008}.  The vertical lines mark the
  critical SMBH mass $M_{\rm tr}$, above which the BH feeding is
  suppressed by strong outflows and jets according to \cite{Yuan_Narayan_2014}.}
\label{fig:M_BH}
\vspace{2\baselineskip}
\end{figure}

We next consider the stable nuclear (sub-pc) region of the accretion
disk, which is embedded by the galactic star-forming disk.  The
properties of this disk are determined primarily by the BH mass and
the gas accretion rate from larger scales ($\ga 1$ pc).
Fig. \ref{fig:M_BH} shows the accretion rate into the nuclear region
for three different Mach numbers $m=0.05$ (blue), $0.1$ (red) and
$m=0.2$ (black).  Up to a turn-over at a critical $M_{\rm BH}$, the
accretion rates in panels (a) and (b) are fit by the single
power-laws,
\begin{equation}
\dot{M}_{\rm BH}\simeq 4.2~m_{0.1}M_{\rm BH,10}^{0.3}~\msunyr,
\label{eq:mdot_1}
\end{equation}
or
\begin{equation}
\frac{\dot{M}_{\rm BH}}{\dot{M}_{\rm Edd}}\simeq 1.8\times 10^{-2}~m_{0.1}M_{\rm BH,10}^{-0.7},
\label{eq:mdot_2}
\end{equation}
where $m_{0.1}\equiv m/0.1$ and $M_{\rm BH,10}\equiv M_{\rm BH}/(10^{10}~\msun)$.
These scaling relations can be explained by an analytical argument in the Appendix.
Assuming a constant radiative efficiency $\eta$, we can integrate
Eq. (\ref{eq:mdot_1}) over the age of the Universe and obtain $M_{\rm
  BH,10}\simeq 7.4~m_{0.1}^{10/7}(1-\eta)^{10/7}$.  This suggests that
SMBHs would not grow above $\sim 10^{11}~\msun$ within the finite age
of the Universe.

\begin{figure*}
\centering
\includegraphics[width=130mm]{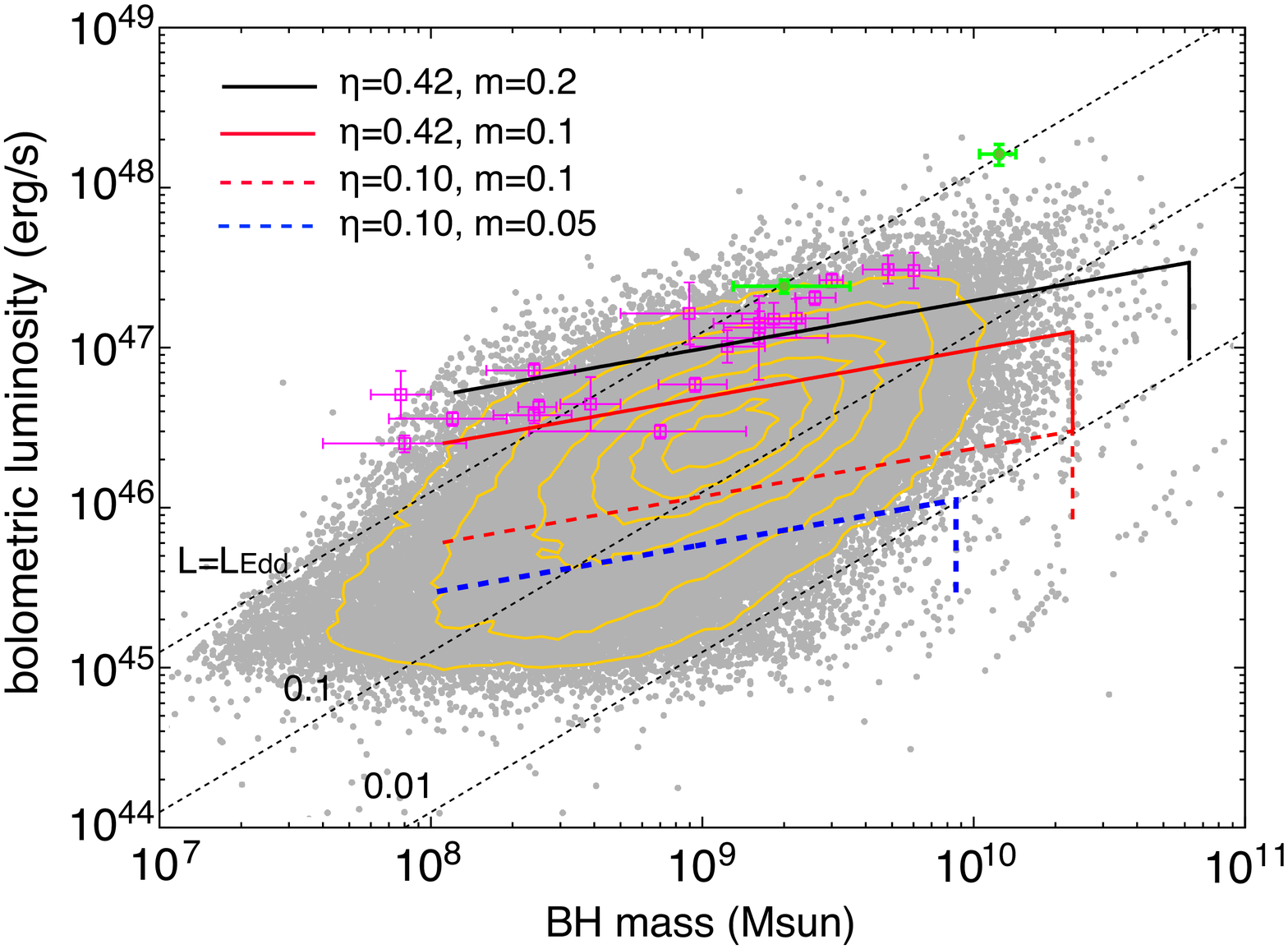}
\caption{Comparison of the predicted $L_{\rm bol}-M_{\rm BH}$ relation
  with observational data.  The data are taken from the AGN/QSOs
  samples in \cite{Shen_2011} for $0<z<5$ (gray dots) and from several
  other studies for $z>5$ (magenta; \citealt{2010AJ....139..906W},
  \citealt{DeRosa_2011} and green; \citealt{Mortlock_2011},
  \citealt{Wu_2015}).  The orange lines shows isodensity contours of
  these samples.  The four thick lines correspond to the $L_{\rm
    bol}-M_{\rm BH}$ relation with different radiative efficiencies $0.1\leq \eta \leq 0.42$
  and Mach numbers $0.05 \leq m \leq 0.2$.  The diagonal dotted lines
  indicate constant Eddington ratios ($L/L_{\rm Edd}$), with values as
  labeled in the bottom left of the figure.  }
\label{fig:M_L}
\vspace{2\baselineskip}
\end{figure*}

The above argument yields a maximum BH mass, which comes close to the
largest observed masses.  Here, we discuss further constraints on the
maximum value, considering properties of accretion flows in the
vicinity of the BH.  In panel (b), the normalized rate for $M_{\rm
  BH}\simeq 10^9~\msun$ is $\mdot \sim 0.1$.  For this value, a
standard geometrically thin, optically thick nuclear disk can form
\citep{Shakura_Sunyaev_1973}.  Through the disk, the BH grows via
accretion at a rate given by Eq. (\ref{eq:mdot_1}).  On the other
hand, the normalized rate decreases with BH mass and reaches the
critical value of $\mdot \la 10^{-2}$, at which the nuclear disk can
not remain thin, because of inefficient radiative cooling
\citep{Narayan_McClintock_2008}.  The inner disk would then likely be
replaced by a radiatively-inefficient ADAF (advection-dominated
accretion flow;
\citealt{Narayan_Yi_1994,Narayan_Yi_1995}).\footnote{\cite{2013ApJ...767..105L}
  discussed a transition to a rotating accretion flow. For
  $\dot{M}/\dot{M}_{\rm Edd}\la 10^{-1.5}$, the rotating flow results
  in a solution with an even lower accretion rate and conical wind
  outflows.}  Adopting the critical rate to be $\mdot=10^{-2}$, we
find that the transition occurs at $M_{\rm BH}\ga M_{\rm tr}=2.3\times
10^{10}m_{0.1}^{10/7}~\msun$ for $0.05\la m \la 0.2$.  We note that
{\it the transition BH mass $M_{\rm tr}$ does not depend on the model
  parameters of the star-forming disk, except on the Mach number $m$},
while the behavior of the accretion rate at $M_{\rm BH}>M_{\rm tr}$
depends on the choices of the other model parameters.

Once the transition to an ADAF occurs, the accretion flow through
the disk becomes hot because of inefficient cooling.  The hot gas near
the BH would launch strong outflows and jets, suppressing the feeding
of the BH \citep[e.g.][]{Blandford_Begelman_1999}.  
The location of the transition radius $R_{\rm tr}$, inside which a
thin disk turns into a hot ADAF, has been discussed by several authors
\citep[e.g.][and references therein]{Yuan_Narayan_2014}.  Although
this location is uncertain, the theoretical models suggest $R_{\rm
  tr}/R_{\rm Sch}\ga 300-10^3~[\dot{M}_{\rm BH}/(10^{-2}\dot{M}_{\rm
    Edd})]^{-q}$ ($q>0$) for $\mdot \la 10^{-2}$.  This value is
consistent with results obtained from fitting the spectra of observed
BH accretion systems ($R_{\rm tr}\sim 100~R_{\rm Sch}$ for $\mdot \sim
10^{-2}$; \citealt{2004ApJ...612..724Y}).  Moreover, numerical
simulations of ADAFs suggest that the gas accretion rate decreases
inward within the transition radius as $\dot{M}_{\rm BH}\propto
(r/R_{\rm tr})^s$
(\citealt{1999MNRAS.303..309I,1999MNRAS.310.1002S,2002ApJ...573..738H,2004ApJ...611..977M},
see also
\citealt{Blandford_Begelman_1999}).  The power-law index is estimated
as $0.4\la s\la 0.6$ at $2\la r/R_{\rm Sch}\la 10^4$, independent of
the strength of viscosity and magnetic field \citep{Yuan_2012a}.  For
a conservative estimate, we set $s=0.3$ and $R_{\rm tr}=100~R_{\rm
  Sch}$ for $\mdot \leq 10^{-2}$.  This reduces the BH feeding rate by
a factor of $(R_{\rm Sch}/R_{\rm tr})^{0.3}\simeq 0.25$ from the
original accretion rate at $R_{\rm tr}$ (Eq. \ref{eq:mdot_1}).  As a
result, the BH growth time is roughly given by $\sim 16~[\dot{M}_{\rm
    BH}/(10^{-2}\dot{M}_{\rm Edd})]^{-1}$ Gyr, and we conclude that
once an SMBH reaches the critical mass of $M_{\rm BH,max}\simeq M_{\rm
  tr} \simeq (0.9-6.2)\times 10^{10}~\msun$, it cannot gain
significant mass within the Hubble time.

\section{Comparison to observations}
\label{sec:BH_lum}

In the TQM05 disk model, the BH feeding rate is a function of SMBH
mass (Eq. \ref{eq:mdot_1}).  We can compare the corresponding predictions
for the $L_{\rm bol}-M_{\rm BH}$ relation (where $L_{\rm bol}$ is the
bolometric luminosity), with observational data. For this comparison,
we use AGN/QSO samples from
\citet{Shen_2011}\footnote{http://das.sdss.org/va/qso$\_$properties$\_$dr7/dr7.htm}
for $0<z<5$ and from \citet{2010AJ....139..906W}, \citet{DeRosa_2011},
\citet{Mortlock_2011} and \citet{Wu_2015} for $z>5$.

For simplicity, we estimate the bolometric luminosity of the nuclear
BH disk assuming a constant radiation efficiency ($L_{\rm bol}=\eta
\dot{M}_{\rm BH}c^2$) as long as the disk is thin,
i.e. $\mdot>10^{-2}$.  The radiative efficiency depends on the BH
spin.  Although we do not have any direct measurements of the SMBH
spin evolution, applying the Paczynski-Soltan
\citep{1982MNRAS.200..115S} argument to the differential quasar
luminosity function, \citet{YuTremaine2002} have inferred typical
radiative efficiencies of $\epsilon\gsim 0.3$ for the brightest
quasars with the most massive SMBHs ($M_{\rm BH}\gsim 10^9{\rm
  M_\odot}$), consistent with rapid spin.  Recently,
\cite{Trakhtenbrot_2014} independently suggested that high-redshift
SMBHs with $\sim 10^{10}~\msun$ have rapid spin with $a\simeq1$, based
on the band luminosities in accretion disk models
\citep[e.g.][]{Davis_Laor_2011}.  Semi-analytical models and numerical
simulations have predicted that a high value of the BH spin is
maintained ($a\simeq 1$) for high-$z$ SMBHs growing via sustained
accretion of cold gas \citep{Volonteri_2007,Dubois_2014}.  Here, we
consider two opposite limits for the efficiency; $\eta=0.1$ that is often used 
and $\eta =0.42$ for an extreme Kerr BH ($a=1$).

In Fig. \ref{fig:M_L}, we show the $L_{\rm bol}-M_{\rm BH}$ relation
predicted for four different combinations of BH spin and Mach number.
As explained in \S\ref{sec:subpc} above, once the BH mass exceeds the
critical value $M_{\rm tr}$, the normalized accretion rate falls below
$\mdot <10^{-2}$, and the BH feeding drops.  Within the range of model
parameters shown in the figure, the maximum BH mass is in good
agreement with the observational data ($M_{\rm BH}\lsim {\rm few}
\times 10^9{\rm M_\odot}$), but favors high values of $a$ and $m$. The
bolometric luminosities are predicted to be between $\sim
10^{45}-10^{47}$ erg s$^{-1}$, in good agreement with the values found
in the AGN/QSO samples. 
Moreover, the slope we predict (${\rm d}\ln L_{\rm bol}/{\rm d}\ln M_{\rm BH}=0.3$)
agrees well with the upper envelope of these samples.
However, the model would require a higher $m$
to reach the luminosities of the rarest bright objects ($\sim$1\% of
all sources) with $\ga 2\times 10^{47}$ erg s$^{-1}$
(e.g. J010013.02+280225.8; \citealt{Wu_2015}).

We briefly note uncertainties of the inferred bolometric luminosities.
The bolometric luminosity is typically estimated from the luminosity
measured in a narrow wave-length range, using a constant conversion
factor based on template spectra
\citep[e.g.][]{1994ApJS...95....1E,2006AJ....131.2766R}.  However,
overestimates of the conventional correction factor from the optical
luminosities have been discussed \citep{Trakhtenbrot_2012}, and
several studies have suggested that the bolometric correction factors
depend on $M_{\rm BH}$ \citep{2008ApJS..176..355K} and increase with
the Eddington ratio $L/L_{\rm Edd}$ \citep{2007MNRAS.381.1235V}.
Thus, this method to estimate $L_{\rm bol}$ would have intrinsic
uncertainties, especially for high-z QSOs.  In addition, beaming could
be present, and produce overestimates of $L_{\rm bol}$ for QSOs with
weak emission lines (e.g. \citealt{Haiman_Cen_2002}).  Since the
fraction of weak-line QSOs is higher at $z\simeq 6$ than at lower
redshift \citep{2014AJ....148...14B}, the bolometric luminosities
could be overestimated for these high-$z$ sources.

\section{Discussion}
\label{sec:discussion}

\subsection{Maximum BH mass of the brightest QSOs}

Among observed SMBHs, the brightest QSOs with $\gsim 2\times 10^{47}$
erg s$^{-1}$, which are inferred to have Eddington ratios near unity
($L\sim L_{\rm Edd}$), would grow at a rate of $\sim \dot{M}_{\rm
  Edd}$.  In the TQM05 model we adopted, this would require a high
radial Mach number ($m\ga 1$).
However, such a large $m$ is unlikely to be realized by global spiral
waves in a marginally stable disk $Q\simeq 1$.  Instead, this rapid
inflow could be triggered by a major galaxy merger, and sustained for
a few dynamical timescales of a few $\times 10^7$ yr
\citep{2010MNRAS.407.1529H,Hopkins_2011}.  After a brief burst phase,
the BH feeding rate would decrease to the value given by
Eq. (\ref{eq:mdot_1}). As long as these major-merger trigged inflows
are sufficiently rare and brief, the SMBH masses will remain limited
by the physics of the star-forming disks, as discussed in
\S\ref{sec:subpc}.

\begin{figure}
\vspace{\baselineskip}
\centering
\includegraphics[width=80mm]{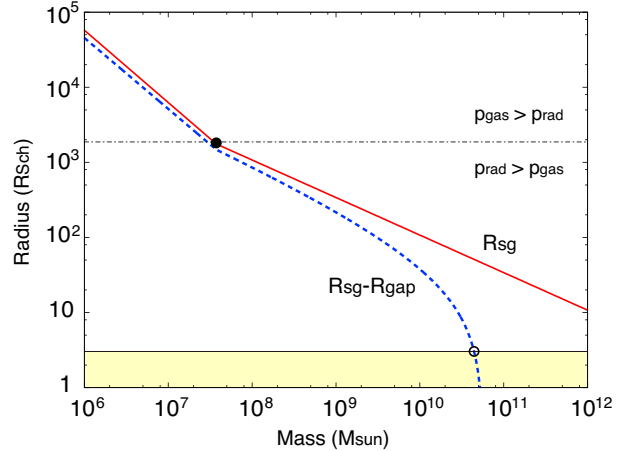}
\caption{The fragmentation radius $R_{\rm sg}$ of a standard
  Shakura-Sunyaev disk in units of $R_{\rm Sch}$ (solid red curve) for
  $\alpha =0.1$ and $\mdot =1$.  The filled circle marks the location
  where $p_{\rm gas} = p_{\rm rad}$ (with radiation pressure
  dominating at higher BH mass).  The dashed curve shows the value of
  ($R_{\rm sg}-R_{\rm gap}$) for $f_{\rm H} = 1.5$, where $R_{\rm
    gap}$ is the radial size of the annular gap cleared by accretion
  onto a clump in circular orbit at $R_{\rm sg}$.  At $M_{\rm BH} =
  4.5\times 10^{10}~\msun$ (open circles), a stable disk cannot exist
  (i.e. $R_{\rm sg}-R_{\rm gap}\approx R_{\rm ISCO}$) and the BH
  feeding would be suppressed.  We set the ISCO radius to $R_{\rm
    ISCO}=3~R_{\rm Sch}$ (horizontal solid line).  }
\label{fig:M_R}
\vspace{2\baselineskip}
\end{figure}

We next argue that BH growth at $M_{\rm BH}\ga 10^{10}~\msun$ would
also be suppressed by fragmentation of the nuclear disk, even at the
higher accretion rates of $\sim \dot{M}_{\rm Edd}$.  In this case, the
disk becomes cold and thin instead of a hot ADAF.  Such a thin disk is
better described by the standard $\alpha$-viscosity prescription
\citep{Shakura_Sunyaev_1973}.  The $\alpha$-disk becomes
self-gravitating and unstable at large radii, where $Q\la 1$,
\begin{equation}
\frac{R_{\rm sg}}{R_{\rm Sch}}=
 \begin{cases}
8.1~\alpha_{0.1}^{14/27} 
\dot{m}_{\rm BH}^{-8/27}
M_{\rm BH,10}^{-26/27}\hspace{5mm}(p_{\rm gas}>p_{\rm rad}),\\
\\
85.3~\alpha_{0.1}^{1/3}
\dot{m}_{\rm BH}^{1/6}
M_{\rm BH,10}^{-1/2}\hspace{11.5mm}(p_{\rm rad}>p_{\rm gas}),
 \end{cases}
\label{eq:R_sg}
\end{equation}
where $\alpha_{0.1}\equiv\alpha/0.1$ and $\dot{m}_{\rm
  BH}\equiv\dot{M}_{\rm BH}/\dot{M}_{\rm Edd}$
\citep{Goodman_Tan_2004}.  The top (bottom) expression is valid when
gas (radiation) pressure dominates.  Fig.~\ref{fig:M_R} shows the
fragmentation radius $R_{\rm sg}$ as a function of the BH mass (solid
curve).  The filled circle marks the location where $p_{\rm
  gas}=p_{\rm rad}$, inside of which radiation pressure dominates
($M_{\rm BH}\ga 4\times 10^7~\msun$).

Gas clumps formed in the unstable region ($r\ga R_{\rm sg}$)
subsequently grow via gas accretion from the ambient disk and the gas
near the clump within $\sim f_{\rm H}R_{\rm H}$ is depleted, where
$R_{\rm H}$ is the clump's Hill radius and $f_{\rm H}\sim O(1)$.
Assuming that the clump grows until a density gap is created, the mass
reaches a substantial fraction of the isolation mass
\citep{Goodman_Tan_2004},
\begin{equation}
M_{\rm c,iso}\simeq 1.3\times 10^9~\alpha_{0.1}^{-1/2}\dot{m}_{\rm BH}^{5/4}M_{\rm BH,10}^{7/4}~f_{\rm H}^{3/2}\msun,
\end{equation}
where the clump location is set to $r=R_{\rm sg}$.
The width of the gap is estimated as 
$R_{\rm gap}\approx f_{\rm H}R_{\rm H}\approx f_{\rm H}R_{\rm sg}
({M_{\rm c,iso}}/{3M_{\rm BH}})^{1/3}$, and thus
\begin{align}
\frac{R_{\rm gap}}{R_{\rm sg}}\approx 0.36~f_{\rm H}^{3/2}\alpha_{0.1}^{-1/6} {\dot{m}_{\rm BH}}^{5/12} M_{\rm BH,10}^{1/4}.
\end{align}
Fig.~\ref{fig:M_R} shows the value of ($R_{\rm sg}-R_{\rm gap}$) for
$f_{\rm H}=1.5$ (dashed blue curve).  A stable disk can exist only
below this line, down to the inner-most stable circular orbit (ISCO),
$R_{\rm ISCO}\simeq 3R_{\rm Sch}$.  The size of the stable region
shrinks with increasing BH mass, and disappears entirely at $M_{\rm
  BH} = 4.5\times 10^{10}~\msun$ (i.e. $R_{\rm sg}-R_{\rm gap}\approx
R_{\rm ISCO}$).  Subsequently, the BH could not be fed via a stable
disk.  Instead, the BH could be fed stars from a nuclear star cluster,
forming by the gravitational collapse of a massive clump at $R_{\rm
  sg}$ with $M_{\rm c,iso}$.  The stellar feeding occurs on the
timescale of $\simeq t_{\rm relax}\ln(2/\theta_{\rm lc})$
\citep[e.g.][]{1976MNRAS.176..633F,1999MNRAS.306...35S}, where $t_{\rm
  relax}$ is the (two-body) relaxation timescale, estimated as
\begin{align}
t_{\rm relax}\simeq \frac{0.34~\sigma_\ast^3 }{G^2 M_{\ast 2}
  \rho_\ast \ln (M_{\rm BH}/M_\ast)}\simeq 6~{\rm Gyr},
\end{align}
\citep{Binney_Tremaine_2008,Kocsis_Tremaine_2011},
where $\sigma_\ast = [ G(M_{\rm BH}+M_{\rm c,iso})/R_{\rm sg} ]^{1/2}$
is the stellar velocity dispersion, $M_{\ast 2}$ is the ratio of the
mean-square stellar mass to the mean stellar mass of the stars, and
$\rho_\ast =3M_{\rm c,iso}/(4\pi R_{\rm sg}^3)$ is the stellar density
of the cluster.  Assuming the Salpeter IMF with $M_{\rm
  min(max)}=1~(100)~\msun$, we obtain $M_{\ast 2} \simeq 11~\msun$.
Since the angular size of the loss cone is estimated as $\theta_{\rm
  lc}=\sqrt{2 R_{\rm Sch}GM_{\rm BH}}/(\sigma_\ast R_{\rm sg})\sim 0.19$ and
$\ln (2/\theta_{\rm lc})\sim 2.4$, the stellar feeding time for $M_{\rm
  BH}>4.5 \times10^{10}~\msun$ exceeds the age of the Universe (at
$z=0$). Therefore, we expect disk fragmentation to suppress BH growth
above this mass (placing the corresponding upper limit of $L\simeq
L_{\rm Edd}\simeq 6\times 10^{48}$ erg s$^{-1}$ on the luminosity).

We note that \citet{King_2016} recently proposed the
existence of an upper limit on the masses of SMBHs, due to
fragmentation of the nuclear disk.  \citet{King_2016} suggests that
the maximum mass is the one for which the fragmentation radius is
located at the ISCO.  
This is very similar to our discussion of the
case of the $\mdot (\simeq 1)$ $\alpha$-disk above. 
The main difference is that \citet{King_2016} adopts a gas-pressure dominated disk, 
though the radiation pressure in fact dominates at $R_{\rm sg}$
for $M_{\rm BH}\ga 4\times 10^7~\msun$
(below dashed-dotted line in Fig.~\ref{fig:M_R}).
\citet{King_2016} argues that a large radiation--pressure dominated disk extending all the way
out to $R_{\rm sg}$ would be thermally unstable and can not form at all.
Then, the BH mass limit is estimated as $\simeq 3\times 10^{10}~\msun$ 
from $R_{\rm sg}\simeq R_{\rm ISCO}$ assuming $p_{\rm gas}>p_{\rm rad}$.
As the implications of this instability are not yet understood,
we here conservatively assumed that a radiation-pressure dominated
disk could still feed the central BH, as long as it is gravitationally
stable.  This, in principle, would greatly increase the fragmentation
radius (see red solve curve in Fig.~\ref{fig:M_R}).  However, we
argued that the large physical size of the clumps in this case
prevents a stable disk from forming all the way down to smaller radii,
comparable to the $R_{\rm sg}$ in the fiducial gas-dominated case (see
dashed blue curve in Fig.~\ref{fig:M_R}). As a result, our main
conclusion agrees with that of \citet{King_2016}.

\begin{figure}
\vspace{\baselineskip}
\centering
\includegraphics[width=80mm]{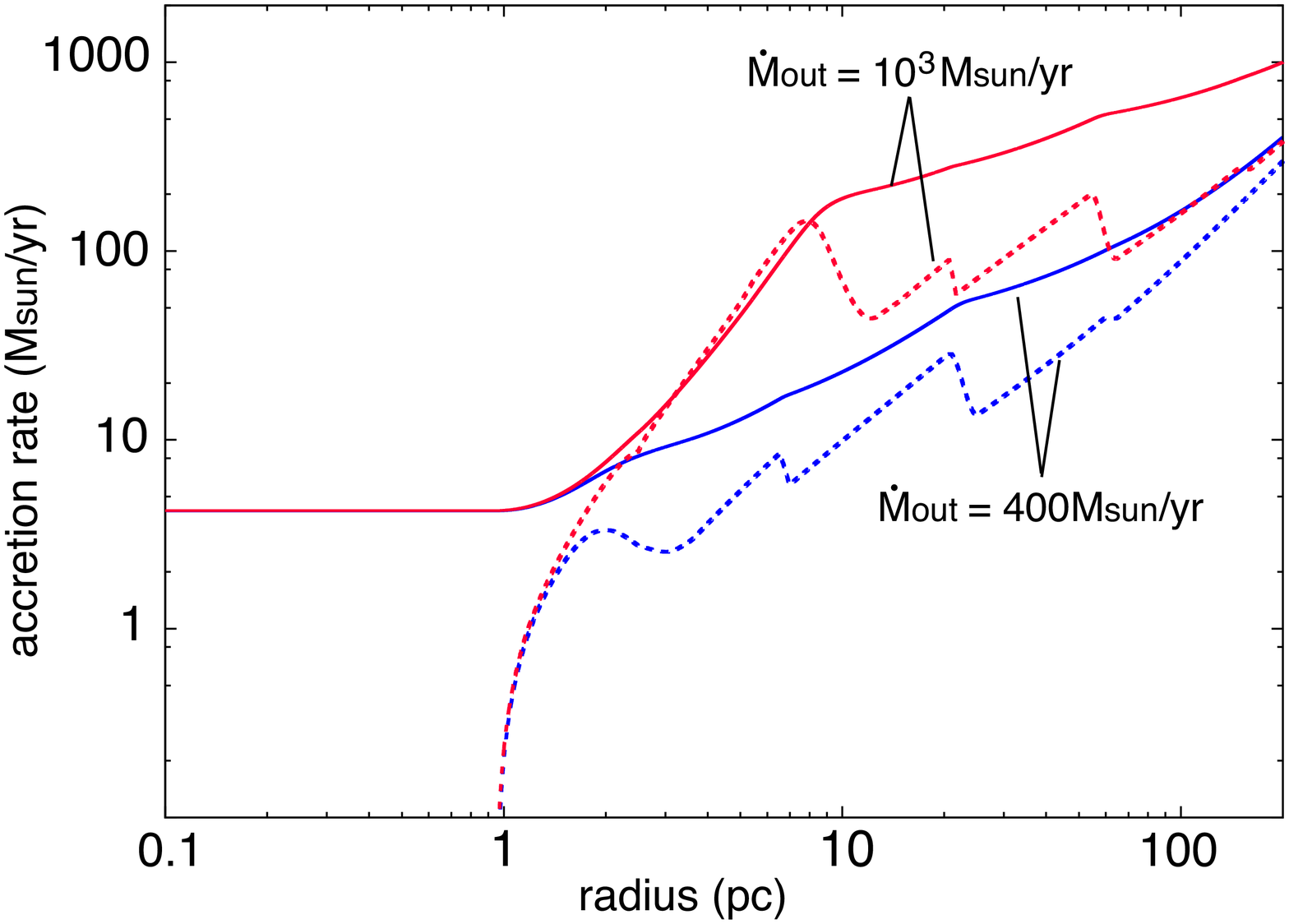}
\caption{The same as Fig. \ref{fig:TQM} but $\dot{M}_{\rm
    out}=400~\msunyr$ (blue) and $10^3~\msunyr$ (red) ($M_{\rm
    BH}=10^{10}~\msun$ and $m=0.1$).  For both the cases, the BH
  feeding rate within $1$ pc is identical.}
\label{fig:SF}
\vspace{2\baselineskip}
\end{figure}

\subsection{$M_{\rm BH}-M_\ast$ relation for the most massive BHs}

In the star-forming disk model, a high accretion rate is required to feed
the central BH (\S\ref{sec:mass_limit} and \S\ref{sec:BH_lum}).  
This fact means that a large amount of stars would
form around the SMBHs.  We briefly discuss the stellar mass of massive
galaxies hosting the most massive BHs with $\sim 10^{10}~\msun$.

Fig. \ref{fig:SF} shows radial profiles of the gas accretion rate and
star formation rate for the two different values of $\dot{M}_{\rm
  out}=400$ and $10^3~\msunyr$ (for $M_{\rm BH}=10^{10}~\msun$ and
$m=0.1$).  We note that $\dot{M}_{\rm out}=400~\msunyr (>\dot{M}_{\rm crit})$ is sufficient
to maintain the universal feeding rate in Eq. (\ref{eq:mdot_1}).
In the case with $\dot{M}_{\rm out}=400~\msunyr$, a population of stars
with total mass $\sim 10^{12}~\msun$ forms within $\sim 200$ pc
in the mass doubling time of the $10^{10}~\msun$ BH ($\sim 2.5$ Gyr).
Note that such a compact star forming region is consistent with observed 
ultra-luminous infrared galaxies where stars form in a few $100$ pc nuclear disk 
at a rate up to several $100~\msunyr$ \citep{2014ApJ...784...70M}.

The most massive elliptical galaxies are as old as $\ga 8$ Gyr
\citep[e.g.][]{2003AJ....125.1882B,2005ApJ...621..673T}.  Thus, we can
observe stars with masses of $<1.1~\msun$, whose lifetimes are longer
than the age of the galaxies (at least $8$ Gyr).  
Although the IMF of stars around the most massive BHs is highly uncertain, 
many authors have discussed the possibility that stars forming in SMBH disks, including
those observed in the Galactic center,  have a top-heavy IMF 
(e.g. \citealt{2006ApJ...643.1011P,Levin_2007,Nayakshin_2007}, 
and see also \citealt{Goodman_Tan_2004}).
Assuming the Salpeter IMF with $M_{\rm min(max)}=1~(100)~\msun$, 4 \%
of the stars in mass live in longer lifetimes of $>8$ Gyr and can be
observed in the most massive elliptical galaxies.  Therefore, we can
estimate the stellar mass surface density as $\sim 3\times
10^{11}~\msun~{\rm kpc}^{-2}$, which is consistent with a maximum
value of dense stellar systems within a factor of three
(\citealt{2010MNRAS.401L..19H}, see also
\citealt{2007ApJ...664..226L}).

\subsection{Super(Hyper)-Eddington growth of intermediate massive BHs}

We briefly mention rapid growth of intermediate massive BHs.
According to Eq. (\ref{eq:mdot_2}), the BH feeding rate in the Eddington units 
$\dot{M}_{\rm BH}/\dot{M}_{\rm Edd}(\propto M_{\rm BH}^{-0.7})$ 
exceeds unity at $M_{\rm BH}\la 3.2\times 10^7~m_{0.1}^{10/7}~\msun$.
In the regime of $\dot{M}_{\rm BH}\ga \dot{M}_{\rm Edd}$, 
the nuclear accretion disk transits to an optically thick ADAF solution, so-called slim disk,
where super-Eddington accretion would be possible
\citep[e.g.][]{1988ApJ...332..646A, 2014ApJ...796..106J,2015MNRAS.447...49S}.
However, radiation heating suppresses gas supply from larger scales,
which results in a lower accretion rate $\sim \dot{M}_{\rm Edd}$
\citep[e.g.][]{2001ApJ...551..131C,2011ApJ...737...26N,2012ApJ...747....9P}.
For intermediate massive BHs with $M_{\rm BH}\la 10^{4}~\msun$,
the BH feeding rate becomes $\dot{M}_{\rm BH}\ga 3000~m_{0.1}L_{\rm Edd}/c^2$,
where hyper-Eddington accretion could be realized unimpeded by radiation feedback,
and the massive BHs would grow rapidly \citep{IHO_2015,Ryu_2016}.

\vspace{\baselineskip}
\section{Summary and Conclusions}
\label{sec:conclusions}

Observations of SMBHs have revealed an upper limit of a few $\times
10^{10}~\msun$ on their mass, in both the local and the early
Universe, nearly independent of redshift.  In this paper, we have
interpreted this to imply that the growth of SMBHs above this mass is
stunted by small-scale physical processes, independent of the
properties of their host galaxies or of cosmology.  The growth of more
massive SMBHs requires a high rate ($\gsim 10^3~\msunyr$) of cold gas
supply from galactic scales into a nuclear region.  We have argued
that even if gas is supplied to the galaxy at such high rates, most of
the gas forms stars at larger radii ($\sim 100$ pc).  Adopting the
model by TQM05 for a star-forming disk, the accretion rate in the
sub-pc nuclear region is reduced to the smaller value of at most $\sim
4~\msunyr (M_{\rm BH}/10^{10}~\msun)^{0.3}$.  This prevents SMBHs from
growing above $\simeq 10^{11}~\msun$ in the age of the Universe.
Furthermore, at this low rate ($\mdot \la 10^{-2}$), the nuclear BH
disk can not maintain a thin structure and changes to a radiatively
inefficient ADAF.  Once this transition occurs, the BH feeding is
further suppressed by strong outflows from hot gas near the BH.  The
maximum mass of SMBHs is given by the critical mass where this
transition occurs, $M_{\rm BH,max}\simeq
(0.9-6.2)\times10^{10}~\msun$, and depends primarily on the angular
momentum transfer efficiency in the galactic disk, and only weakly on
other properties of the host galaxy.

Although this model gives a compelling explanation for the observed
maximum SMBH masses, it underpredicts, by a factor of few, the highest
observed quasar luminosities.  These rare high-luminosity objects
would require a high (near-Eddington) accretion rate, but we have
argued that they do not significantly add to the SMBH masses, because
these bursts may correspond to brief episodes following major mergers,
and because we find that self-gravity prevents a stable accretion disk
from forming for $M_{\rm BH}>4.5 \times10^{10}~\msun$ even in this
high-$\mdot$ regime.

Finally, if the explanation proposed here is correct, it requires that
stars forming in disks around the most massive SMBHs have a top-heavy
IMF, in order to avoid over-producing the masses of compact nuclear
star-clusters in massive elliptical galaxies.  This is consistent
with theoretical expectations.

\section*{Acknowledgements}
We thank Jeremiah Ostriker, Yuri Levin, Nicholas Stone, Benny
Trakhtenbrot, Kazumi Kashiyama, Shy Genel, Kohei Ichikawa and Jia Liu
for fruitful discussions.
This work is partially supported by Simons
Foundation through the Simons Society of Fellows (KI), and by NASA
grants NNX11AE05G and NNX15AB19G (ZH).

\appendix
\section{Analytical derivations of the scaling relations}

We here give derivations of the scaling relations of 
$\dot{\Sigma}_\ast \propto \Sigma_{\rm g}/\kappa$ (\S\ref{sec:TQM}) and 
$\dot{M}_{\rm BH} \propto mM_{\rm BH}^{1/3}$ (\S\ref{sec:subpc}), 
and an analytical expression of $\dot{M}_{\rm crit}$ (\S\ref{sec:TQM}).
These arguments are based on TQM05 (see their \S2 and Appendix A).

In a star-burst disk, we assume that the accretion disk is marginally stable against the self-gravity 
($Q\simeq 1$ or $\Sigma_{\rm g}\propto c_{\rm s}\Omega$)
and is a hydrostatic equilibrium state to the vertical direction, the total pressure is given by
\begin{equation}
p=\rho h^2\Omega^2=\Sigma_{\rm g}c_{\rm s}\Omega \propto  \Sigma_{\rm g}^2.
\label{eq:1}
\end{equation}
For a radiation-pressure dominant ($p\simeq p_{\rm rad}\propto T^4$) and optically thick ($\tau\gg1$) disk,
the pressure is expressed as $p\propto \tau  \dot{\Sigma}_\ast $ (see Eq. \ref{eq:prad}).
Combining these relations with $\tau \simeq \kappa \Sigma_{\rm g}$, we can obtain two relations
\begin{equation}
\dot{\Sigma}_\ast \propto \Sigma_{\rm g}/\kappa,
\label{eq:3}
\end{equation}
\begin{equation}
T\propto \Sigma_{\rm g}^{1/2}.
\label{eq:4}
\end{equation}
As we discussed in \S\ref{sec:TQM}, the star formation rate increases at radii, where 
dust opacity decrease by sublimation (e.g. $T_{\rm dust,sub}\simeq 1000~\K$),
to maintain the marginally-stable disk structure.

Next, we derive the relation of the BH feeding rate $\dot{M}_{\rm BH}$ with
the BH mass and the Mach number of the radial velocity ($m=v_{\rm r}/c_{\rm s}$).
As the gas temperature in the disk increases inward and reaches $T_{\rm dust,sub}(\simeq 1000~\K)$,
the opacity rapidly decreases because of dust sublimation 
($\kappa \propto T^{\beta}$ at $T\ga T_{\rm dust,sub}$, where $\beta < -20$).
In this opacity gap, higher star formation rates are required to support the disk
in the vertical direction via radiation pressure (see Eq. \ref{eq:3}).
Because of the gas consumption, the gas accretion rate decreases inward inside the opacity gap,
where timescales of the star formation $t_\ast \equiv \Sigma_{\rm g}/\dot{\Sigma}_\ast$ and 
the radial advection $t_{\rm adv}\equiv r/v_r$ are balanced.
These timescales are estimated as 
\begin{equation}
t_\ast \propto \kappa \propto T^{\beta}, 
\end{equation}
\begin{equation}
t_{\rm adv}
\propto \frac{r\Omega}{\Sigma_{\rm g}m}
\propto \frac{r\Omega}{T^2m},
\end{equation}
where we use Eqs. (\ref{eq:3}) and (\ref{eq:4}).
Thus, the condition where $t_\ast \simeq t_{\rm adv}$ gives us a relation of
\begin{align}
T\propto \left(\frac{rm^2}{M_{\rm BH}}\right)^{-1/(4+2\beta)},
\label{eq:6}
\end{align}
which means that $T\simeq T_{\rm dust,sub}$ is kept inside the opacity gap.
Since the accretion and the star formation are balanced ($\dot{M}\sim r^2\dot{\Sigma}_\ast $), 
we obtain a relation from Eqs. (\ref{eq:3}) and (\ref{eq:6})
\begin{align}
\dot{M}
\propto r^2T^{2-\beta}
\propto r^{\frac{6+5\beta}{4+2\beta}}
M_{\rm BH}^{\frac{2-\beta}{4+2\beta}}
m^{\frac{\beta-2}{2+\beta}}
~\longrightarrow ~
r^{5/2} M_{\rm BH}^{-1/2} m
~~~~~{\rm for}~\beta \rightarrow -\infty.
\label{eq:7}
\end{align}
The accretion rate decreases approximately following $\dot{M}\propto r^{5/2}$ in the opacity gap,
where the temperature does not change but the density increase toward the center.
As a result, the gas pressure dominates the radiation pressure eventually, and thus
star formation becomes less important as a energy source to support the disk structure.
We estimate the characteristic radius $R_{\rm gas}$ within which $p_{\rm gas}> p_{\rm rad}$.
From the equation of continuity (Eq. \ref{eq:Sigma}), we estimate
\begin{equation}
p_{\rm gas}
\simeq \Sigma_{\rm g}c_{\rm s}\Omega \sim \frac{\dot{M}\Omega}{rm}.
\end{equation}
Since $p_{\rm gas}\propto \Omega^2 T$ and $p_{\rm rad}\propto T^4$, 
the condition of $p_{\rm gas}\simeq p_{\rm rad}$ gives
\begin{equation}
p_{\rm rad}\propto \left(\frac{M}{r^3}\right)^{4/3},
\end{equation}
and thus we obtain 
\begin{equation}
R_{\rm gas}\propto \dot{M}^{-2/3}M_{\rm BH}^{5/9}m^{2/3}.
\label{eq:10}
\end{equation}
At $r<R_{\rm gas}$, the star formation rate becomes below the gas accretion rate and thus $\dot{M}(r)\simeq {\rm const}$,
which is the BH feeding rate $\dot{M}_{\rm BH}$.
Substituting Eq. (\ref{eq:10}) into Eq. (\ref{eq:7}), we find the relation 
\begin{equation}
\dot{M}_{\rm BH} \propto mM_{\rm BH}^{1/3},
\label{eq:11}
\end{equation}
which is in good agreement with Eqs. (\ref{eq:mdot_1}) and (\ref{eq:mdot_2}).
Combining Eqs. (\ref{eq:10}) and (\ref{eq:11}), we obtain $R_{\rm gas}\simeq 1.4~M_{\rm BH,10}^{7/9}~{\rm pc}$.
Note that viscous heating is still subdominant at $r=R_{\rm gas}$, 
but stabilizes the disk at $r<R_{\rm gas}$, where the Toomre parameter exceeds unity ($Q>1$).

Finally, we estimate the critical gas accretion rate $\dot{M}_{\rm crit}$ at a large radius $R_{\rm out}$.
For $\dot{M}_{\rm out}>\dot{M}_{\rm crit}$, the gas accretion rate is high enough 
to maintain the universal BH feeding rate (Eqs. \ref{eq:mdot_1} and \ref{eq:11}).
Otherwise, the gas in the disk is depleted due to efficient star formation at $\sim R_{\rm out}$ and 
thus the BH feeding rate becomes much lower than the universal value. 
Since the dust opacity is given by $\kappa=\kappa_0T^2$ at the large radius, 
where the gas temperature is $\la 100~\K$, the star formation timescale is $t_\ast \propto \epsilon \kappa_0T^2$.
Thus, the accretion rate at $R_{\rm out}$ required to feed the BH at the universal rate (Eq. \ref{eq:mdot_1}) is given by
$t_\ast \ga t_{\rm adv}(\simeq \Sigma_{\rm g}R_{\rm out}^2/\dot{M}_{\rm out}\propto T^2R_{\rm out}^2/\dot{M}_{\rm out})$, 
that is,
\begin{equation}
\dot{M}_{\rm out}\ga \dot{M}_{\rm crit} \simeq 280~\msunyr
\left(\frac{R_{\rm out}}{200~{\rm pc}}\right)^2
\left(\frac{\epsilon}{10^{-3}}\right)^{-1}
\left(\frac{\kappa_0}{2.4\times 10^{-4}~{\rm cm^2~g^{-1}~K^{-2}}}\right)^{-1},
\label{eq:12}
\end{equation}
(see also Eq. 44 in TQM05).

\bibliography{ref}

\end{document}